\documentclass[prl,twocolumn,floatfix,superscriptaddress]{revtex4}
\usepackage{graphicx}
\usepackage{amssymb}

\begin{document}
\title{Correlation induced 
  resonances in transport through coupled quantum dots}
\author{V.~Meden}
\affiliation{Institut f\"ur Theoretische  Physik, Universit\"at G\"ottingen, 
D-37077 G\"ottingen, Germany}
\author{F.~Marquardt}
\affiliation{Sektion Physik, Universit\"at M\"unchen, D-80333 
M\"unchen, Germany}

\begin{abstract}
We investigate the effect of local electron correlations 
on transport through parallel quantum dots. 
The linear conductance as a function of gate voltage is strongly 
affected by the interplay of the interaction $U$ and quantum 
interference. We find a pair of novel correlation induced resonances 
separated by an energy scale that depends exponentially on $U$.
The effect is robust against a small detuning
of the dot energy levels and occurs for arbitrary generic tunnel
couplings. It should be observable in experiments on the basis of
presently existing double-dot setups.
\end{abstract}
\maketitle     

The theoretical and experimental research 
on electronic transport through ultrasmall quantum 
dots has become a very active field. 
Various fundamental physical phenomena such as quantum
interference,\cite{HeiblumExps} Coulomb blockade,\cite{Sohn} 
and the Kondo effect,\cite{Glazman,Goldhaber}
strongly affect the transport properties. 
Currently, the focus is shifting towards few-electron 
double-dot
structures, 
that are studied as artificial 
molecules, interferometers, and for charge- and spin-based quantum
computing.\cite{Loss,Florian,Holleitner1,Chen,Sigrist,Holleitner2,Petta1,Koppens,Petta2} 
Theoretical investigations into the role of correlations 
in systems of two or more coupled quantum 
dots are still at the beginning and much remains to be explored. 
In parallel quantum dots connected to common 
leads new physics is to be expected due to the interplay of
correlations and quantum interference, an issue that is of interest 
also from a broader perspective.

\begin{figure}[t]
\begin{center}
\includegraphics[width=.27\textwidth,clip]{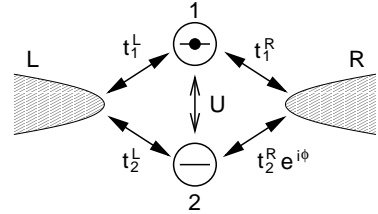} 
\end{center}

\vspace{-0.7cm}

\caption[]{System of coupled quantum dots with common leads.\label{fig1}}
\end{figure}

In the present Letter, we 
investigate this problem for a specific model of two  parallel 
quantum dots coupled by an electron interaction $U$ (see
Fig.~\ref{fig1}). The effect of external electrostatic potentials on
two-path interference was studied earlier (magneto-electric
Aharonov-Bohm effect).\cite{Nazarov} We here investigate the role of 
interaction induced potentials. 
This is especially relevant for molecular transport, as the interaction
induced potentials will be far larger than
the external potentials which are difficult to apply to such small
structures.
We study how the linear conductance 
$G$ as a function of gate voltage $V_g$ changes with increasing $U$. 
Considering the entire parameter space we find a very rich generic 
behavior and predict the appearance of novel correlation induced 
resonances (CIRs) if $U$ is larger than a critical interaction $U_c$.  
The effect is robust: It appears for almost arbitrary combinations 
of the four tunnel couplings and also remains visible for 
a small detuning of the dot level energies. The separation of the resonances 
in gate voltage defines an energy scale that depends exponentially 
on $U$ and on a combination of the tunnel couplings. 
It is argued that this new correlation effect is unrelated to Kondo
physics. We employ a 
powerful new method, the functional renormalization group (fRG), to
efficiently obtain both numerical and analytical results for this 
many-body problem, and we have confirmed all the essential features 
of our results using the numerical renormalization group (NRG). 
Double-dot geometries that could form the basis to verify 
our predictions have been experimentally 
realized in Refs.~\onlinecite{Holleitner1,Chen,Sigrist,Holleitner2}. 
Our model is equally of relevance for 
transport through two nearly degenerate levels of a single dot, a 
subject that has attracted much attention recently in attempts to 
understand the puzzling behavior of the transmission phase\cite{HeiblumExps}.

We study a system of two quantum dots $j=1,2$ each having a single
level $\varepsilon_j$ as sketched in Fig.~\ref{fig1}.
The dots are coupled by a Coulomb interaction 
$U \geq 0$ and are connected 
to two common leads $l=L,R$ via tunnel barriers $t_j^l$. 
The dot Hamiltonian is 
$H_{\rm dot} = \sum_{j} \varepsilon_j d_j^\dag d_j + U (n_1-1/2) 
(n_2-1/2)$ and the dot-lead coupling is given by 
$H_T=- \sum_{j,l} (t_j^l c^\dag_{0,l} 
d_j + \mbox{H.c.})$, where $c^\dag_{0,l}$ denotes the creation 
operator at the end of the semi-infinite lead $l$.  
The leads are modeled by $H_l = -t \sum_{m=0}^{\infty} (c^\dag_{m,l}
c_{m+1,l} + \mbox{H.c.} )$. 
The energy scale of the dot level broadening is given by $\Gamma_j^l =  
\pi |t_j^l|^2 \rho_l $, where $\rho_l$ denotes the local density 
of states at the end of lead $l$. As usual we later take $\rho_l$ to be energy 
independent (wide band limit). 
The ring structure is pierced by a magnetic flux $\phi$ that 
we take into account by multiplying $t_2^R$ by a phase 
factor $e^{i \phi}$ ($\phi= 2 \pi$ corresponds to the flux 
quantum $h c/e$). For symmetry reasons one only has to consider $0
\leq \phi \leq \pi$. The levels are shifted by a common 
gate voltage $V_g$ such that $\varepsilon_1=V_g+\delta$ and 
$\varepsilon_2=V_g$, where $\delta$ denotes a detuning. 
For $\delta=0$ due to the shift of $n_j$ in 
$H_{\rm dot}$, $V_g=0$ corresponds to half filled dots. 
We neglect the spin of the electron and thus suppress the spin
Kondo effect. Experimentally, the contribution of
spin physics may be excluded by applying a sufficiently strong magnetic
field.\cite{Holleitner2} We focus on temperature $T=0$. 
The spectral properties\cite{Boese1}  and level 
occupancies\cite{Sindel1,Koenig} of this model were investigated earlier. 

To compute $G$ and, in addition, the level occupancies $\left< n_j\right>$ 
we mainly use a recently developed fRG scheme.\cite{VM1} The starting 
point is an exact hierarchy of 
differential flow equations for the real-space self-energy matrix 
$\Sigma^{\Lambda}$ and higher
order vertex functions, where $\Lambda \in (\infty,0]$ 
denotes an infrared energy cutoff which is the flow parameter.
We truncate the hierarchy by neglecting the flow of the two-particle
vertex only considering $\Sigma^{\Lambda}$, which is then
energy independent. This approximation and variants of it 
were successfully used to study a variety of 
transport problems through quasi one-dimensional wires of 
correlated electrons 
(Tomonaga-Luttinger liquids).\cite{VM2}   
In addition, spectral and transport properties of locally correlated
systems were investigated using the
fRG.\cite{Hedden} Although the guiding principle behind the
above approximation scheme is perturbation theory, it was shown that
depending on the problem studied, the fRG procedure correctly leads to 
power-laws with $U$ dependent exponents (Tomonaga-Luttinger
liquids)\cite{VM2} as well as exponential 
dependencies on $U$ (Kondo effect).\cite{Hedden}  

For the double-dot $\Sigma^\Lambda$ 
is a  $2 \times 2$ matrix in the dot label $j$. The diagonal parts
are real and $V_j^\Lambda = \Sigma^{\Lambda}_{j,j}+V_g$ can be considered 
as effective dot level positions. The (for arbitrary $\phi$) complex 
off-diagonal contribution $t_d^\Lambda = - \Sigma^{\Lambda}_{1,2}$ is a  
hopping between the two dot states generated by the interaction.      
The flow equations are 
\begin{eqnarray}
\label{floweq}
\partial_\Lambda V_j^{\Lambda} & = & - \frac{U}{2 \pi} \sum_{\omega = \pm
  \Lambda} {\mathcal G}^\Lambda_{\bar j, \bar j}(i \omega) \; ,\nonumber \\
\partial_\Lambda t_d^{\Lambda} & = & - \frac{U}{2 \pi} \sum_{\omega = \pm
  \Lambda} {\mathcal G}^\Lambda_{1, 2}(i\omega) \; ,
\end{eqnarray}
with  $\bar j$ being the complement of $j$. 
The Green function is ${\mathcal G}^\Lambda(i\omega) =  
\left[i\omega - h^\Lambda(i\omega)
\right]^{-1}$  with 
\begin{eqnarray}
\label{effh}
h^\Lambda(i\omega) = \left( \begin{array}{cc} 
V_1^\Lambda - i \Gamma_1  \, \mbox{sgn}(\omega)& 
- t_d^\Lambda - i \gamma \, \mbox{sgn}(\omega)\\
- \left( t_d^\Lambda \right)^\ast - i \gamma^\ast \, \mbox{sgn}(\omega) &
V_2^\Lambda - i \Gamma_2 \, \mbox{sgn}(\omega) 
  \end{array} \right) 
\end{eqnarray}
and $\Gamma_j = \sum_l \Gamma_j^l$, $\gamma =  
\sqrt{\Gamma_1^L \Gamma_2^L} + e^{i \phi} \sqrt{\Gamma_1^R \Gamma_2^R}$.   
The initial conditions are $V_1^{\Lambda= \infty} = V_g + \delta$,
$V_2^{\Lambda= \infty} = V_g$, and $t_d^{\Lambda= \infty}=0$.
To obtain an
approximation for $\Sigma$ and the one-particle Green function 
(using the Dyson equation) one has to solve the system (\ref{floweq})
of four real 
coupled differential equations. This can easily be done numerically
and for a specific class of parameters also analytically. The occupancies
$\left< n_j \right>$ can directly be calculated from the Green
function. For the present
problem the same holds  for $G$ as current vertex 
corrections vanish. One can easily derive a 
lengthy expression for $G$ in terms of the parameters 
$\Gamma_j^l,\phi$ and the renormalized level positions  
$V_j=V_j^{\Lambda=0}$ and hopping $t_d = t_d^{\Lambda=0}$ 
not presented here. $V_j$ and $t_d$ depend on $\Gamma_j^l,\phi$ 
as well as on $V_g$, $\delta$, and $U$. 

We use the  NRG as a nonperturbative 
method to confirm the essential validity of the physics discovered within
the fRG. We allow arbitrary dot-lead
couplings by employing the Kubo-formula for the current, which yields
the frequency-dependent conductance 
\begin{equation}
{\rm Re}\,
G(\omega)=\frac{\pi}{\omega}\sum_{f}\left|I_{0,f}\right|^{2}
\delta(E_{f}-E_{0}-\omega)\, , \;\;\; \omega >0 \, .
\end{equation}
The current operator $I=e(\dot{N}_{R}-\dot{N}_{L})/2$ 
(with $N_{l}=\sum_{m=0}^{\infty}c_{m,l}^{\dagger}c_{m,l}$
and $\dot{N}_{l}=i[H,N_{l}]$) is expressed in terms of lead- and
dot-operators. Its matrix elements $I_{0,f}$ (with $0$ indicating the
ground state and $f$ the excited states; $E_0$ and $E_f$ are the
respective energies) are evaluated in the 
NRG basis and the $\omega\rightarrow0$ limit then provides the 
conductance $G$.\cite{Sindel2}

We first consider the case of degenerate levels with $\delta=0$ and
later discuss how the results change for $\delta>0$.
The behavior of $G(V_g,U)$ can be cast in four classes that 
can already be identified at $U=0$. (i) If two or more of the 
$\Gamma_j^l$ are 0 such that no closed path between the left and right 
lead exists $G(V_g,U)\equiv 0$. (ii) $\Gamma_1^L \Gamma_1^R = \Gamma_2^L 
\Gamma_2^R$ and $\phi=\pi$: In this case one can introduce new 
fermionic dot states such that one only couples to the left lead and 
the other only to the right lead, implying $G(V_g,U)\equiv 0$. We note 
that in this case (and only in this case) a conserved pseudo-spin 
variable (left/right) exists.\cite{Boese1}  
(iii) A nonvanishing (but nongeneric) conductance is found for
$\Gamma_1^L  \Gamma_2^R =  \Gamma_1^R  \Gamma_2^L \neq 0$ and
$\phi=0$. In this case $G(V_g,U=0)$ is given by a Lorentzian
centered around 0. The $U>0$ dependence of $G(V_g,U)$ can most 
easily be studied in the {\it exactly solvable} 
case of equal $\Gamma_j^l$.\cite{Boese1} 
It is characterized by two Coulomb blockade peaks located at 
$\approx \pm U/2$. (iv) For all other $\Gamma_j^l,\phi$, that is 
for generic parameters on which we focus in the following, 
the peak in $G(V_g,U=0)$ at $V_g=0$ [as in case
(iii)] is replaced by a dip with $G(V_g,U=0)=0$.  For equal 
$\Gamma_j^l$, $\phi=0$, $\delta>0$, and $U=0$ the appearance of 
such dips was discussed earlier and explained as a destructive 
interference between path traversing dots 1 and 2 
respectively.\cite{Silva} In the limit of a strong asymmetry of the 
transmission probability via dots 1 and 2, e.g.~for $\Gamma_1^l 
\ll \Gamma_2^l$, the dip can be viewed as a Fano anti-resonance
resulting from the interference of a resonant path and a path with
energy independent transmission: 
For energies at which the transmission via dot 1 shows a resonance, 
the transmission via dot 2 can be regarded as constant.

\begin{figure}[t]
\begin{center}
\includegraphics[width=.34\textwidth,clip]{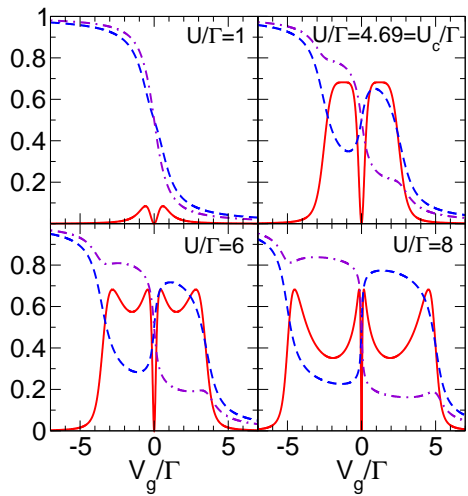} 
\end{center}

\vspace{-0.7cm}

\caption[]{(Color online) Generic results 
for $G(V_g)/(e^2/h)$ (solid lines), $\left< n_1 \right>$ (dashed lines),
and $\left< n_2 \right>$ (dashed-dotted lines) at different $U$ obtained 
from the fRG with $\Gamma_1^L=0.27 \Gamma$, $\Gamma_1^R=0.33 \Gamma$, 
$\Gamma_2^L= 0.16 \Gamma$, $\Gamma_2^R=0.24 \Gamma$, $\phi=\pi$, and 
$\delta=0$. The two novel correlation induced resonances are visible 
in the lower panels (large $U$), near $V_g=0$.
\label{fig2}}
\end{figure}

Fig.~\ref{fig2} shows the generic evolution of $G(V_g)$ 
for increasing $U$ at $\delta=0$. Because of particle-hole
symmetry $G$ is  symmetric around $V_g=0$. Energies are given 
in units of $\Gamma= \sum_j \Gamma_j$. 
Increasing $U$ the height of the two peaks resulting from the dip 
at $V_g=0$ increases and the maximum flattens. At a critical 
$U=U_c(\{\Gamma_j^l\},\phi)$ each of the peaks splits into two. 
For the present example the fRG approximation is 
$U_c/\Gamma \approx 4.69$. Further increasing $U$ the two outer most 
peaks move towards larger $|V_g|$ and become the Coulomb blockade 
peaks located at $V_g \approx \pm U/2$. 
The other two peaks at $\pm V_{\rm CIR}$ are the {\it novel CIRs,} 
where $V_{\rm CIR} >0$ decreases with increasing $U$. 
Associated with $G(V_g=0)=0$ at $U=0$ is a jump of the transmission 
phase by $\pi$. As the phase evolves continuously with $U$ and 
particle-hole symmetry holds for any $U$, the $\pi$-phase-jump 
and thus $G(V_g=0)=0$ must remain for all $U$.
For $U> U_c$ the height of all four peaks is equal 
to $h_{\rm max}(\{\Gamma_j^l\},\phi) \leq e^2/h$ and does not 
change with $U$.

For equal $\Gamma_j^l$, $\phi=\pi$, and $\delta=0$ [case (ii) above]  
the conserved pseudo-spin leads to orbital Kondo physics.\cite{Boese1} 
Remnants of this effect were found in the 
vicinity of this parameter point but die out quickly away from 
it. As our correlation effect appears 
generically, in particular for parameters far away from the Kondo 
point it is apparently unrelated to Kondo physics. 

In addition to $G(V_g)$ in Fig.~\ref{fig2} we present the dot occupancies. 
For small $U$ the $V_g$ dependence of the 
occupancies of dots 1 and 2 is monotonic. In the opposite limit of 
large $U$, $\left< n_j \right>$ depends non-monotonically 
on $V_g$.\cite{Sindel1,Koenig} Starting at negative $V_g$ the level 
that is coupled more strongly (here $j=1$) is depopulated at the 
first Coulomb blockade peak, while the occupancy of the other 
level stays close to 1. Close to $V_g=0$  we find an 
inversion of the population followed by 
another depopulation of the more strongly coupled dot across 
the second Coulomb blockade peak at 
$V_g \approx U/2$. Note that the non-monotonic behavior of the 
$\left< n_j \right>$ sets in for interactions smaller than $U_c$ and
is thus not directly related to the appearance of the CIRs. This is
consistent with the observation that in contrast to the CIRs (see below) 
the non-monotonicity of $\left< n_j \right>$ can already be observed 
within a self-consistent Hartree approximation.\cite{Sindel1}     
 
\begin{figure}[t]
\begin{center}
\includegraphics[width=.48\textwidth,clip]{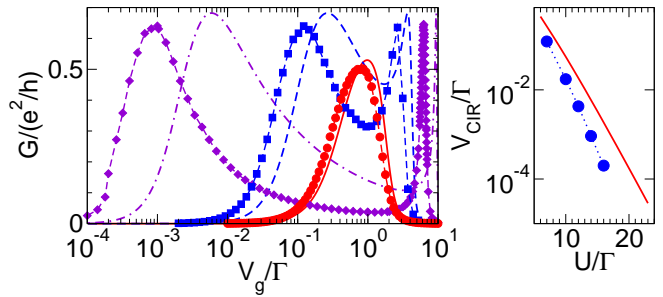} 
\end{center}

\vspace{-0.7cm}

\caption[]{(Color online) Comparison of fRG (lines) and NRG
 (symbols) results for  the same parameters as 
in Fig.~\ref{fig2}. {\it Left:} $G(V_g)$ for different $U$ 
($U=3.5 \Gamma$: solid line and circles; $U=7 \Gamma$: dashed line and
squares; $U=14 \Gamma$: dashed-dotted line and diamonds). 
 {\it Right:} The resonance position $V_{\rm CIR}$ 
as a function of $U$. 
\label{fig3}}
\end{figure}

The left part of Fig.~\ref{fig3} shows a comparison of fRG and 
NRG data for $G$ at $V_g>0$ and for different $U$. 
To clearly resolve the CIRs we use a logarithmic scale. 
The NRG data show {\it all} the features discussed in connection with 
Fig.~\ref{fig2}.  
In particular, for increasing $U > U_c$, $V_{\rm CIR}$ becomes 
small quickly.  
The right part of Fig.~\ref{fig3} 
shows the $U$ dependence of $V_{\rm CIR}$ extracted from
the numerical data. For $U$ sufficiently larger than $U_c$, 
\begin{eqnarray}
V_{\rm CIR}/\Gamma \propto \exp{\left[-
    C\left(\left\{\Gamma_j^l\right\},\phi\right) U/\Gamma\right]} \, ,   
\label{centralresult}
\end{eqnarray}  
with $C>0$. 
By construction the fRG based approximation scheme 
works particularly well for small to intermediate $U$. 
At larger $U$ and for all $\Gamma_j^l,\phi$ we tested the fRG 
overestimates the positions of the Coulomb blockade peaks and the
CIRs. As will be shown in an upcoming publication this can 
systematically be improved using a more elaborate fRG truncation 
scheme. For a specific class of $\Gamma_j^l,\phi$ 
we next {\it analytically} confirm the exponential dependence 
of $V_{\rm CIR}$ on $U$ and derive an explicit expression for 
$C$ using the fRG. In the most general case the dependence 
of $C$ (and $U_c$) on $\Gamma_j^l$ is complex and requires
further investigation. Roughly speaking $C$ increases ($U_c$
decreases) with increasing asymmetry of the $\Gamma_j^l$ [see also 
Eq.~(\ref{explicitexp}) below].
For fixed $\Gamma_j^l$ and 
increasing $0 \leq \phi \leq \pi$, $C$ decreases while $U_c$ 
increases. 

\begin{figure}[t]
\begin{center}
\includegraphics[width=.38\textwidth,clip]{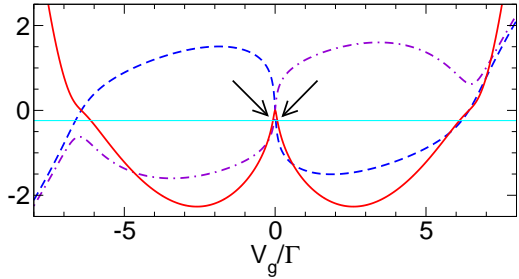} 
\end{center}

\vspace{-0.7cm}

\caption[]{(Color online) Renormalized dot level positions $V_1/\Gamma$
  (dashed line) and $V_2/\Gamma$ (dashed-dotted line). 
  The thick solid line is the
  product $V_1 V_2/\Gamma^2$. The thin solid line lies at $-4
  \Gamma_1^L \Gamma_2^L/\Gamma^2$ and its intersections with the
  thick solid line determine the peak positions of $G$. The two crossings
  indicated by the arrows are at $V_g = \pm V_{\rm CIR}$. 
  The parameters are $U=10 \Gamma$, 
  $\Gamma_1^L= \Gamma_1^R =0.3 \Gamma$,  $\Gamma_2^L= \Gamma_2^R =0.2
  \Gamma$, $\phi=\pi$, and $\delta=0$ ($U_c/\Gamma \approx 5.05$ from fRG). 
\label{fig4}}
\end{figure}

We now consider $\Gamma_1^L = \Gamma_1^R$, $\Gamma_2^L = \Gamma_2^R$,
but $\Gamma_1^L \neq \Gamma_2^L$, and $\phi=\pi$. 
In this case $\gamma=0$ in Eq.~(\ref{effh}) and the off-diagonal
elements of the Green function ${\mathcal G}^\Lambda$ are proportional to
$t_d^{\Lambda}$. Initially (at $\Lambda=\infty$) $t_d^{\Lambda}$ 
vanishes and it will thus remain zero during the fRG flow which leads 
to a simplification of the flow equations (\ref{floweq}).
For small $V_j^{\Lambda}$, that is small $V_g$, these equations 
can be solved analytically and in the limit 
$U \gg |\Gamma_1^L - \Gamma_2^L|$ we obtain 
\begin{eqnarray}
\label{explicitexp}
V_{\rm CIR} /\Gamma \propto \exp{\left[ - \frac{U}{2 \pi} \, 
\frac{\ln{(\Gamma_1^L/\Gamma_2^L)}}{\Gamma_1^L - \Gamma_2^L} \right]}
\; .
\end{eqnarray}  
In Fig.~\ref{fig4} we show the renormalized level positions $V_j$ as a
function of $V_g$ for fixed $U>U_c$. 
For $|V_g| \gg U/2$, $V_1$ and
$V_2$ become equal and are given by $V_g - \mbox{sgn}(V_g) \, U/2$. 
For $-U/2 \lessapprox V_g < 0$ the position of the level that is
coupled more weakly (here $j=2$) is smaller than the chemical
potential $\mu=0$, while the other level has energy larger than
$\mu$. For $0< V_g \lessapprox U/2$  the role of the two levels is
interchanged. This explains the observed 
$V_g$ dependence of the $\left< n_j \right>$ discussed in connection with
Fig.~\ref{fig2}. 
Peaks in $G$ are found at $V_1  V_2  = - 4  \Gamma_1^L \Gamma_2^L$. 
In particular, the crossings of $V_1 V_2$ (thick solid
line) and $-4 \Gamma_1^L \Gamma_2^L$ (thin solid line) 
at $\pm V_{\rm CIR}$ occur 
because $V_1$ and $V_2$ continuously go through $0$ at
$V_g=0$, one coming from above the other from below. This is not 
the case if the self-consistent Hartree-Fock approximation 
is used to compute the effective level positions. Within this 
approach $V_1 V_2$ discontinuously jumps at $V_g=0$ leading to only 
two crossings at $V_g \approx \pm U/2$ associated 
with the Coulomb blockade peaks. This shows that similar to the Kondo 
effect more sophisticated methods than the Hartree-Fock approximation 
are required to describe local correlation effects. The fRG is such a 
method (see also Ref.~\onlinecite{Hedden}). 

\begin{figure}[t]
\begin{center}
\includegraphics[width=.40\textwidth,clip]{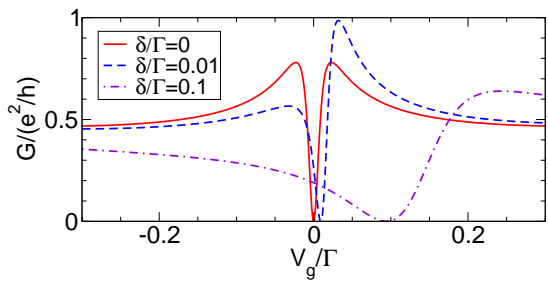} 
\end{center}

\vspace{-0.7cm}

\caption[]{(Color online) Dependence of $G(V_g)$ on the bare level
  splitting $\delta$. The parameters are 
  $U=3 \Gamma$, $\Gamma_1^L=0.7 \Gamma$, 
  $\Gamma_1^R=0.2 \Gamma$, $\Gamma_2^L= 0.02 \Gamma$, 
  $\Gamma_2^R=0.08 \Gamma$, and $\phi=0$ ($U_c/\Gamma \approx 0.842$ 
from fRG). 
\label{fig5}}
\end{figure}

We now investigate the stability of the CIRs in the presence of an initial
level splitting $\delta>0$. In Fig.~\ref{fig5} we show fRG 
data for $G(V_g)$ with different $\delta$ at small $|V_g|$. 
For $\delta > \Gamma$ and all parameter sets we
studied $G$ only shows the two Coulomb blockade peaks at $V_g \approx
-U/2-\delta$ and $V_g \approx U/2$. The way 
how this large $\delta$ limit is reached depends on the specific 
choice of parameters but common to {\it all} cases, remnants 
of the CIRs were clearly observable for 
$\delta \ll \Gamma$, as exemplified in
Fig.~\ref{fig5}. We found a similar $\delta$ dependence using NRG. 
A direct hopping between 
the two dots induces a level splitting if the noninteracting 
dot Hamiltonian is diagonalized and has thus an effect similar 
to that of $\delta > 0$.

To summarize, we found that the transport properties of  
a double-dot interferometer are strongly affected by electron correlations.
We discovered a novel pair of correlation induced resonances 
that should be measurable in double-dots of appropriate 
geometry\cite{Holleitner1,Chen,Sigrist,Holleitner2} 
in the presence of strong Zeeman splitting. Varying the tunnel barriers
(and thus $U/\Gamma$) it should be possible to study the entire
scenario discussed above. 
Apparently this correlation effect is unrelated to both 
spin\cite{Glazman} and orbital\cite{Boese1} Kondo physics.
Rather, it follows from the interplay
of local correlations and quantum interference. It is thus likely 
that similar effects will appear in transport through 
more complex systems as e.g.~ring-like molecules studied in the
context of molecular electronics.    
Besides revealing interesting new physics, 
we showed that the fRG method is a very promising tool 
to investigate problems with local Coulomb correlations.
In comparison to NRG the fRG is far superior in terms of the
numerical effort required, e.g.~enabling efficient analysis of parameter
dependencies. Furthermore, the fRG can easily be extended to more complex 
systems with local electron correlations. 

We thank J.~v.~Delft, T.~Enss, W.~Hofstetter, Th.~Pruschke, H.~Schoeller, 
K.~Sch\"onhammer, A.~Sedeki, and A.~Weichselbaum for valuable 
discussions and in particular M.~Sindel for supplying us with his NRG code.  
V.M.~is grateful to the Deutsche Forschungsgemeinschaft (SFB 602) 
for support.


\begin{thebibliography}{*}

\bibitem{HeiblumExps} R.~Schuster {\it et al.}, Nature {\bf 385}, 417 (1997);
M. Avinun-Khalish {\it et al.}, Nature {\bf 436}, 529 (2005), and Refs. therein.\bibitem{Sohn} {\it Mesoscopic Electron Transport,} edited by
  L.L.~Sohn, L.P.~Kouwenhoven, and G.~Sch\"on (Kluwer, Dodrecht, 1997).
\bibitem{Glazman} L.~Glazman and M.~Raikh, JETP Lett. {\bf 47}, 452 (1988); 
T.~Ng and P.~Lee, Phys.~Rev.~Lett.~{\bf 61}, 1768 (1988).
\bibitem{Goldhaber} D.~Goldhaber-Gordon {\it et al.,} Nature {\bf
    391}, 156 (1998); W.~van der Wiel {\it et al.,} Science {\bf
    289}, 2105 (2000). 
\bibitem{Loss} D.~Loss and D.P.~DiVincenzo, Phys.~Rev.~A {\bf 57},
  120 (1998); D.~Loss and E.V.~Sukhorukov, Phys.~Rev.~Lett.~{\bf
    84}, 1035  (2000).
\bibitem{Florian} F.~Marquardt and C.~Bruder, Phys.~Rev.~B~{\bf 68}, 
195305 (2003).
\bibitem{Holleitner1} A.~W.~Holleitner {\it et al.,} 
Phys.~Rev.~Lett. {\bf 87}, 256802 (2001); Science {\bf 297}, 70 (2002).
\bibitem{Chen} J.C.~Chen, A.M.~Chang, and M.R.~Melloch, Phys.~Rev.~Lett.~{\bf
  92}, 176801 (2004).
\bibitem{Sigrist} M.~Sigrist {\it et al.,} Phys.~Rev.~Lett.~{\bf
  93}, 066802 (2004).
\bibitem{Holleitner2} A.~W.~Holleitner {\it et al.,} 
Phys.~Rev.~B {\bf 70}, 075204 (2004). 
\bibitem{Petta1} J.R.~Petta {\it et al.,} Phys.~Rev.~Lett.~{\bf 93}, 
186802 (2004). 
\bibitem{Koppens} F.H.L.~Koppens {\it et al.,} Science {\bf 309}, 
1346 (2005).
\bibitem{Petta2} J.R.~Petta {\it et al.,} Science {\bf 309}, 
2180 (2005).
\bibitem{Nazarov} Yu.V.~Nazarov, Phys.~Rev.~B {\bf 47}, 2768 (1993); 
  A.~van Oudenaarden, M.H.~Devoret, Yu.V.~Nazarov, and J.E.~Mooij, 
  Nature {\bf 391}, 768 (1998); W.G.~van der Wiel {\it et
    al.,} Phys.~Rev.~B {\bf 67}, 033307 (2003). 
\bibitem{Boese1} D.~Boese, W.~Hofstetter, and H.~Schoeller, 
Phys.~Rev.~B {\bf 64}, 125309
  (2001). 
\bibitem{Sindel1} M.~Sindel, A.~Silva, Y.~Oreg, and J.~von Delft,
Phys.~Rev.~B {\bf 72}, 125316
(2005).
\bibitem{Koenig} J.~K\"onig and Y.~Gefen, Phys.~Rev.~B {\bf 71},
  201308(R) (2005).
\bibitem{VM1}
V.~Meden, W.~Metzner, U.~Schollw\"ock, and K.~Sch\"on\-hammer, 
J.~Low Temp.~Phys.~{\bf 126}, 1147 (2002).
\bibitem{VM2} V.\ Meden {\it et al.,} Europhys.~Lett.~{\bf 64}, 769
  (2003);  T.~Enss {\it et al.,} 
Phys.~Rev.~B {\bf 71}, 155401 (2005); X.~Barnab\'e-Th\'eriault,
A.~Sedeki, V.~Meden, and K.~Sch\"onhammer,  
Phys.~Rev.~Lett.~{\bf 94}, 136405 (2005). 
\bibitem{Hedden} R.~Hedden, V.~Meden, Th.~Pruschke, and
  K.~Sch\"on\-ham\-mer,  J.~Phys.: Condens.~Matter
  {\bf 16}, 5279 (2004); S.~Andergassen, T.~Enss, and V.~Meden,
cond-mat/0509576.
\bibitem{Sindel2} M.~Sindel, W.~Hofstetter, J.~von Delft, and
  M.~Kindermann, Phys.~Rev.~Lett.~{\bf 94}, 196602 (2005). 
\bibitem{Silva} A.~Silva, Y.~Oreg, and Y.~Gefen, Phys.~Rev.~B {\bf 66}, 195316
  (2002). 
\end{thebibliography}
\end{document}